\documentclass[usenatbib]{mn2e}
\usepackage{mathtools}
\usepackage{graphicx}
\usepackage{epstopdf} 
\usepackage{natbib}
\usepackage{verbatim}

 \usepackage{times}
\def\lsim{\mathrel{\rlap{\lower4pt\hbox{\hskip1pt$\sim$}}
    \raise1pt\hbox{$<$}}}                
\def\gsim{\mathrel{\rlap{\lower4pt\hbox{\hskip1pt$\sim$}}
    \raise1pt\hbox{$>$}}}                

\begin{document}

\title[]
{Spectroscopic detection of CIV$\lambda$1548 in a galaxy at $z=7.045$: Implications for the ionizing spectra of reionization-era galaxies}

\author[Stark et al.] 
{Daniel P. Stark$^{1}$\thanks{dpstark@email.arizona.edu}, 
Gregory Walth$^{1}$,
St\'{e}phane Charlot$^{2}$,
Benjamin Cl\'{e}ment$^{3}$,
Anna Feltre$^2$,\newauthor
Julia Gutkin$^2$, 
Johan Richard$^3$,
Ramesh Mainali$^1$, 
Brant Robertson$^{1}$,
Brian Siana$^{4}$, \newauthor 
Mengtao Tang$^1$ \&
Matthew Schenker$^5$
\vspace{0.1in}\\
$^{1}$ Steward Observatory, University of Arizona, 933 N Cherry Ave, Tucson, AZ 85721 USA \\  
$^{2}$ Sorbonne Universit\'{e}s, UPMC-CNRS, UMR7095, Institut d'Astrophysique de Paris, F-75014 Paris, France \\
$^{3}$ Centre de Recherche Astrophysique de Lyon, Universite Lyon 1, 9 Avenue Charles Andre, 69561,  France  \\
$^{4}$ Department of Physics \& Astronomy, University of California, Riverside, CA 92507 USA \\
$^{5}$ PDT Partners, LLC, 1745 Broadway, NY, NY 10019 USA   \\
}

\pagerange{\pageref{firstpage}--\pageref{lastpage}} \pubyear{2015}

\hsize=6truein
\maketitle

\label{firstpage}
\begin{abstract}

We present Keck/MOSFIRE observations of UV metal lines in four bright (H=23.9-25.4) gravitationally-lensed $z\simeq 6-8$ galaxies behind the cluster Abell 1703.     The spectrum of A1703-zd6,  a highly-magnified star forming galaxy with a Ly$\alpha$ redshift of $z=7.045$, reveals a confident (S/N=5.1) detection of the nebular 
CIV$\lambda$1548 emission line (unresolved with FWHM$<$125 km s$^{-1}$).   UV metal lines are not detected in the three other galaxies.  At $z\simeq 2-3$, nebular CIV emission is observed in just 1\% of UV-selected galaxies.   The presence of strong CIV emission in one of the small sample of galaxies targeted in this paper may indicate hard ionizing spectra are more common at $z\simeq 7$.    The total estimated equivalent width of the CIV doublet ($\rm{W}_{\rm{CIV}} \simeq 38$~\AA) and CIV/Ly$\alpha$ flux ratio ($f_{\rm{CIV}}$/f$_{\rm{Ly\alpha}} \simeq 0.3$) are comparable to measurements of narrow-lined AGNs.  Photoionization models show that the nebular CIV line can 
also be reproduced by  a young stellar population, with very hot metal poor stars dominating the photon flux responsible for triply ionizing carbon.  Regardless of the origin of the CIV, we show that the ionizing spectrum of A1703-zd6  is different from that of typical galaxies at $z\simeq 2$, producing more H ionizing photons per unit 1500~\AA\ luminosity ($\rm{log (\xi_{ion}/erg^{-1}}$ Hz)=25.68) and a larger flux density at 30-50 eV.   If such extreme radiation fields are typical in UV-selected systems at $z\gsim 7$, it would indicate that reionization-era galaxies are more efficient ionizing agents than previously thought.    Alternatively, we suggest that the small sample of Ly$\alpha$ emitters at $z\gsim 7$ 
may trace a rare population with intense radiation fields capable of ionizing their surrounding hydrogen distribution. Additional constraints on high ionization emission 
lines in galaxies with and without Ly$\alpha$ detections will help clarify whether hard ionizing spectra are 
common in the reionization era.

\end{abstract}

\begin{keywords}
cosmology: observations - galaxies: evolution - galaxies: formation - galaxies: high-redshift
\end{keywords}

\section{Introduction}

Our understanding of galaxy growth in the first  billion years  of cosmic time  
has developed rapidly in the last five years following a series of deep imaging 
campaigns with the infrared channel of the Wide Field Camera 3 (WFC3/IR)  onboard the {\it Hubble Space Telescope} (HST).
Deep WFC3/IR exposures have delivered  more than $\sim 1500$ galaxies photometrically-selected to lie between $\simeq 0.5-1$ Gyr after the Big Bang (e.g. \citealt{Bouwens2014b}) and the first 
small samples of galaxies within the first 0.5 Gyr of cosmic time (e.g., \citealt{Zheng2012,Ellis2013, Coe2013,Atek2014,Zitrin2014,Oesch2014}).

These studies demonstrate that the $z\gsim 6$ galaxy population  is different from  
well-studied samples at $z\simeq 2-3$.   The UV luminosities, star formation rates, 
and stellar masses  tend to be lower at $z\simeq 6$ (e.g. \citealt{Smit2012,McLure2013,Schenker2013a,Bouwens2014a,Salmon2014,Duncan2014,
Grazian2014}),
the sizes are  smaller (e.g., \citealt{Oesch2010,Ono2013}; c.f., \citealt{Curtislake2014}), and the UV 
continuum colors  are bluer (e.g., \citealt{Wilkins2011,Finkelstein2012,Rogers2013,Bouwens2014a}).    Specific star formation rates at $z\gsim 6$ are large, 
indicating a rapidly growing young stellar population (e.g., \citealt{Stark2013a,Gonzalez2014,Salmon2014}).

The emission line properties of early galaxies also appear different.   Large equivalent width 
Ly$\alpha$ emission is more common among $z\simeq 6$ galaxies than it is in similar 
systems at $z\simeq 3$ (e.g., \citealt{Stark2011,curtislake2012}).  
The strongest rest-frame optical lines ([OIII], H$\alpha$) are more difficult to characterize since they  
are situated at $3-5 \mu$m, where thermal emission from the atmosphere impedes detection with 
ground-based facilities.  Nevertheless progress has been achieved by isolating galaxies at redshifts in which the [OIII] or H$\alpha$ line contaminates the {Spitzer}/IRAC broadband filters.   In the last several 
years, this technique has been used to characterize the equivalent width distribution of H$\alpha$ in $3.8<z<5.0$ galaxies (\citealt{Shim2011,Stark2013a}) and [OIII]+H$\beta$ in $z\simeq 6.6-6.9$ \citep{Smit2014a,Smit2014b} and $z\simeq 8$ \citep{Labbe2013} galaxies, revealing that typical 
[OIII] and H$\alpha$ equivalent widths of $z\simeq 4-7$ galaxies are significantly larger than among similar systems at $z\simeq 2$.    The population of extreme emission line galaxies (EW = 500-1000~\AA), relatively rare among $z\simeq 1-2$ galaxies (e.g., Atek et al. 2011, van der Wel et al. 2011), 
appears to be ubiquitous at $z\simeq 7$ \citep{Smit2014a, Smit2014b}. 

The common presence of  extreme optical line emission holds  clues as to the physical nature 
of $z\gsim 7$ galaxies.   Recent work has demonstrated that optical line ratios among $z\simeq 2$ galaxies require an ionizing radiation field that is harder than in $z\simeq 0$ galaxies 
(e.g., \citealt{Steidel2014}), consistent with a blackbody with mean 
effective temperatures of 50,000-60,000 K.    The increased incidence of extreme 
line emitters at $z\simeq 7$ with respect to $z\simeq 2$ may suggest that the net ionizing field 
is powered by even hotter stars at $z\simeq 7$.  Such a finding would indicate that galaxies are very efficient ionizing agents throughout the tail end of the reionization era and would have 
important implications for the nature of the massive stars  within $z\simeq 7$ galaxies.

Further progress in our understanding of the radiation field of early galaxies will only come from detailed spectroscopy.   Prior to JWST, spectroscopic observations of $z\gsim 7$ galaxies will be limited to the rest-frame far-ultraviolet (FUV)  window.   The FUV contains several emission lines with higher ionization potential than in the rest-frame optical (i.e., CIV, He II), providing a valuable probe of the ionizing spectrum.    
While nebular emission from these species is rarely seen among luminous star forming galaxies, 
they have been identified more commonly in metal poor galaxies with low masses \citep{Erb2010,Christensen2012,Stark2014a}.   Such systems also commonly show strong emission 
from the OIII]$\lambda\lambda$1660,1666 and CIII]$\lambda\lambda$1907,1909 intercombination lines \citep{Garnett1995,Garnett1997,Garnett1999,Erb2010,Bayliss2013,James2014,Stark2014a}

\begin{figure*}
\begin{center}
\includegraphics[trim=0.15cm 0cm 1.0cm 0cm,width=0.95\textwidth]{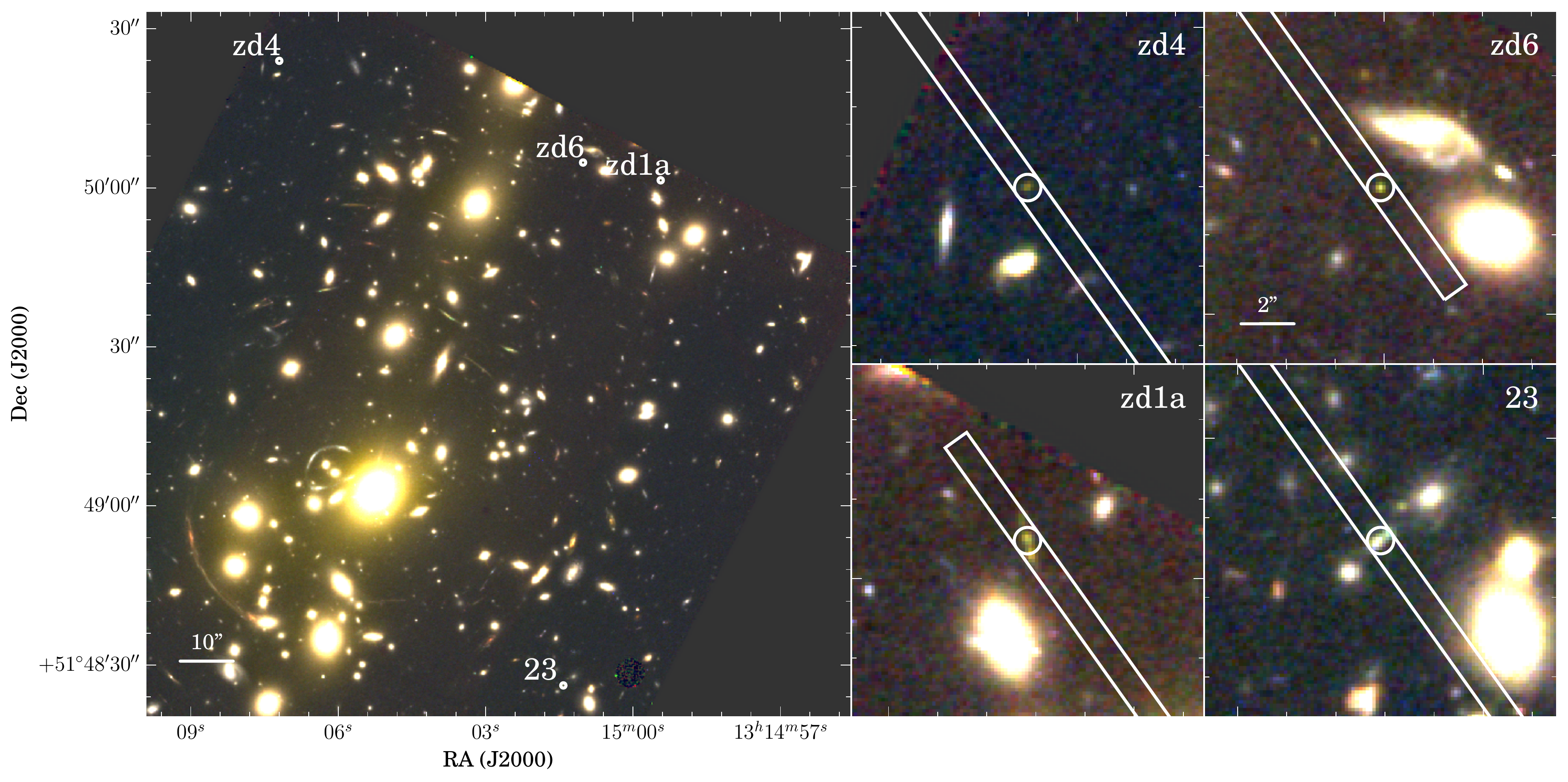}
\caption{Overview of the Keck/MOSFIRE J-band observations of four 
gravitationally-lensed galaxies at $5.8<z<7.0$ in 
the field of Abell 1703.   Positions of the galaxies are overlaid on the HST color image (z$_{850}$, J$_{125}$, H$_{160}$-bands) of the cluster.  
The right panel shows a zoomed in view.    Information on the galaxies and MOSFIRE 
spectra is presented in \S3.
}
\label{fig:hst}
\end{center}
\end{figure*}

Little is currently known about whether FUV spectra of 
galaxies in the reionization era.  But with the discovery of an increasing number of bright $z\gsim 6$ 
galaxies (e.g., \citealt{Bradley2013, Bowler2014}), this is now beginning to change.    Recently 
\citet{Stark2014b} reported   tentative CIII] detections in two galaxies with previously confirmed redshifts ($z=6.029$ and $z=7.213$) from Ly$\alpha$.   Here we build on this progress with an exploration of 
the strength of high ionization emission features (CIV$\lambda\lambda$1548,1551 and He II$\lambda$1640) in a small sample of $z\simeq 7$ UV-selected galaxies using the MOSFIRE \citep{Mclean2012} spectrograph on the Keck I telescope.
We report  the detection of strong nebular  CIV$\lambda$1548 emission in the spectrum of A1703-zd6, a spectroscopically 
confirmed galaxy at $z=7.045$.   The  total CIV equivalent width  is greater than 
in existing samples of $z\simeq 2$ metal poor star forming galaxies \citep{Christensen2012,Stark2014a} but is 
similar to that seen in narrow-lined AGNs \citep{Hainline2011,Alexandroff2013}, pointing to a hard radiation 
field in one of the most distant known galaxies.   

Other galaxies on the MOSFIRE mask include a  spectroscopically confirmed Ly$\alpha$ emitting galaxy at $z=5.828$ \citep{Richard2009} and A1703-zd1a, a photometrically-selected galaxy thought to lie in the range $z\simeq 6.6-6.9$ with {\it Spitzer}/IRAC colours 
(\citealt{Bradley2012, Smit2014a}) indicating extreme [OIII]+H$\beta$ emission in the rest-frame optical, and a bright $z$-band dropout (A1703-zd4 in \citet{Bradley2012}) with photometry suggesting a redshift in the range $z=7.0-9.3$.    No strong emission lines are seen in any of these systems.

The plan of the paper is as follows.   In \S2 we describe the MOSFIRE observations.  The spectra are discussed in \S3, and the photoionization modelling procedure is described in \S4.    We explore implications of our findings 
in \S5 and summarize the results in \S6.
We adopt a $\Lambda$-dominated, flat Universe
with $\Omega_{\Lambda}=0.7$, $\Omega_{M}=0.3$ and
$\rm{H_{0}}=70\,\rm{h_{70}}~{\rm km\,s}^{-1}\,{\rm Mpc}^{-1}$. All
magnitudes in this paper are quoted in the AB system \citep{Oke1983}.

\section{Keck/MOSFIRE observations}
\begin{table}
\begin{tabular}{llcccc}
\hline 
Source  & z$_{\rm{spec}}$ & z$_{\rm{phot}}$ &  H$_{160}$ &  UV lines  targeted   & Ref \\  \hline  \hline
A1703-zd1a &    \ldots & 6.6-6.9 & 23.9 & CIV, He II, OIII]  & [1]   \\ 
A1703-zd4 &   \ldots  & 7.0-9.3  & 25.4 &   Ly$\alpha$, CIV$^\dagger$  & [1] \\ 
A1703-zd6 &  7.043 & \ldots & 25.9 &  CIV, He II, OIII]  & [1] \\ 
A1703-23 &   5.828 & \ldots & 23.8 &  CIII]  &[2] \\ 
   \end{tabular}
\caption{High redshift galaxies targeted with our Keck/MOSFIRE observations.   Details 
on observations and reduction are presented in \S2.  The final column provides the reference to the article where the galaxy was first discussed in the literature.   The  magnitude of 
A1703-23 is from the H-band filter on Subaru/MOIRCS, while the others are from WFC3/IR imaging 
in the H$_{160}$ filter.     For A1703-zd4, the $\dagger$ symbol marks the fact that we are 
sensitive to CIV over only a portion of the redshift range suggested by the photometry.   References: [1]  
\citet{Bradley2012}; [2]  \citet{Richard2009}.   }
\end{table}

Near-infrared spectroscopic observations of Abell 1703 were carried out
with MOSFIRE (McLean et al. 2010, 2012) on the Keck-I Telescope on April
11, 2014 UT.  The spectroscopic observations were taken using the YJ
grating with the J-band filter, which has a resolution of R=3318 and covers
a wavelength range of $\lambda$=1.15-1.35$\micron$.  A mask was created for
Abell 1703 with 1\farcs0 width slits.  Individual exposures were 120 seconds
with two position dithers of 3\farcs0 having a total integration time of 2.6
hours.  The average seeing throughout the observation was 0\farcs80 (FWHM).

The spectra were reduced using the MOSFIRE Data Reduction Pipeline (DRP).
The MOSFIRE DRP performs the standard NIR spectroscopic reduction;
flat-fielding, wavelength calibration, sky-subtraction, and Cosmic Ray
removal to produce 2D spectra.  The flux calibration was performed using
a star placed in the mask of the science field with HST ACS/WFC3
photometry.  The flux from the star was scaled by the HST photometry and
the spectral slope was corrected by fitting a power law between photometric
bands.  The 1D-spectra were extracted with 1\farcs1, 1\farcs4 and 1\farcs8
(6, 8, 10 pixels) apertures.

\section{Results}

In the following, we discuss the physical properties of each of the four galaxies that 
form the basis of this study and report the detections and non-detections arising 
from the MOSFIRE spectra. 

\subsection{A1703-zd6}
A1703-zd6 is a bright (H=25.9) $z$-band dropout first identified in \citet{Bradley2012}.  
A spectroscopic redshift ($z=7.045$) was achieved via detection of Ly$\alpha$ at 
$9780$~\AA\ \citep{Schenker2012}.    The absolute UV magnitude is found to be  M$_{\rm{UV}}=-19.3$ after 
correcting for the source magnification ($\mu=5.2$).

The MOSFIRE J-band spectrum of A1703-zd6 covers 1.1530 to 1.3519 $\mu$m providing 
spectral coverage between 1433~\AA\ and 1680~\AA\ in the rest-frame, enabling constraints on the strength of NIV], CIV, He II, and OIII].  
The Ly$\alpha$ redshift (Schenker et al. 2012) allows us to predict the window over which 
these lines will be located.   In doing so, we must account for the velocity offset between 
Ly$\alpha$ and the other FUV  lines.   We expect the NIV], CIII], and OIII] doublets to trace the systemic 
redshift (Vanzella et al. 2010; Stark et al. 
2014a) and Ly$\alpha$ to be redshifted between 0 and 450 km s$^{-1}$ with respect 
to the systemic redshift.   The CIV$\lambda$1549 doublet is  a resonant line and may also appear redshifted with respect to the other FUV metal lines.
 
The CIV doublet  is easily  resolved by MOSFIRE at $z\simeq 7$.    If CIV  
traces gas at the same redshift as Ly$\alpha$   ($\rm{z=7.045\pm0.003}$), we would 
expect CIV$\lambda$1548 to be located between 1.2450 and 1.2460$\mu$m  and CIV$\lambda$1550 between 1.2471 and 1.2481$\mu$m.  As can be seen in the 
2D spectrum shown in Figure 2, CIV$\lambda$1548 is confidently detected in the window 
defined by Ly$\alpha$ at 1.2458$\mu$m.   CIV$\lambda$1550   is also likely detected 
at 1.2474$\mu$m.   A skyline redward of the line makes determination of the exact centroid 
difficult.  The CIV$\lambda$1548 line 
flux is $4.1\pm0.8$ $\times$10$^{-18}$ erg cm$^{-2}$ s$^{-1}$.  Determination of the CIV$\lambda$1550 flux is more difficult because of the neighbouring sky line.   We measure 
a line flux of $3.8\pm0.8$$\times$10$^{-18}$ erg cm$^{-2}$ s$^{-1}$ blueward of the skyline. 
Taking into account the seeing and the fraction of the galaxy 
covered by the MOSFIRE slit, we estimate that an aperture correction of 1.2$\times$ must be applied to the line flux for accurate equivalent width measurements.
Applying this correction, we derive rest-frame equivalent widths of $19.9\pm4.2$~\AA\ and $18.1\pm 4.3$~\AA\ for the 
CIV$\lambda$1548 and CIV$\lambda$1550 components, respectively. 
  The observed spectral FWHM of CIV$\lambda$1548 (5.2 \AA) is identical to the FWHM of nearby sky lines  indicating 
that the line is narrow and unresolved with a velocity FWHM of $\lsim 125$ $\rm{km~s^{-1}}$.   
  
To predict the wavelengths of the OIII]$\lambda\lambda$1660,1666 doublet, we consider Ly$\alpha$ velocity offsets ($\Delta v_{\rm{Ly\alpha}}$) between 0 and $\rm{450~km~s^{-1}}$, consistent with observations of $z\simeq 2$ galaxies \citep{Tapken2007,Steidel2010}.   Under these assumptions, the OIII]$\lambda$1660 line will lie between 1.3341 $\mu$m ($\Delta v_{\rm{Ly\alpha}}=450$ km s$^{-1}$) and 1.3361$\mu$m ($\Delta v_{\rm{Ly\alpha}}=0$ 
km s$^{-1}$).  An emission feature is visible in this spectral window 
 at 1.3358~$\mu$m (Figure 2).   We tentatively identify this line as OIII]$\lambda$1660.
After accounting for the aperture correction, the measured line flux 
($1.8\pm0.7$ $\times$10$^{-18}$ 
$\rm{erg~cm^{-2}~s^{-1}}$) implies a rest-frame equivalent width of $9.8\pm 3.9$~\AA.   

Since CIV and Ly$\alpha$ are both resonant transitions, OIII]$\lambda$1660 provides the only constraint on the systemic redshift.   From the peak flux of emission feature, we estimate a redshift of $z_{\rm{sys}}=7.0433$.   The Ly$\alpha$ velocity offset that would be implied by this tentative 
detection, $\Delta v_{\rm{Ly\alpha}}$ = 60 km s$^{-1}$, suggests Ly$\alpha$ is emerging 
close to systemic redshift.    A more robust detection is required to verify the Ly$\alpha$ velocity offset.   But 
we note that this measurement would be  consistent with both  Ly$\alpha$ emitting galaxies 
at $z\simeq 2$ \citep{Tapken2007,Mclinden2011,Hashimoto2013}.  Such small velocity 
offsets appear to be  more common at higher redshifts \citep{Schenker2013b,Stark2014b}, 
 consistent with the rising fraction of Ly$\alpha$ emitting galaxies with redshift over 
$3<z<6$ \citep{Stark2011}. 
As discussed in \citet{Choudhury2014}, existence of small Ly$\alpha$ velocity offsets at $z\gsim 6$ reduces the IGM neutral hydrogen fraction required to explain the rapid attenuation of Ly$\alpha$ over $6<z<7$.

At $z=7.0433$, the OIII]$\lambda$1666 is located on top of a strong skyline at 1.340$\mu$m (Figure 2) and is not detected in the MOSFIRE spectrum.     
We also do not detect He II$\lambda$1640 or NIV]$\lambda\lambda$1483,1487.     Flux and equivalent width 
limits (2$\sigma$) are provided in Table 2.   Non-detection of He II and NIV] is consistent with 
the emission line spectra of  $z\simeq 1.5-3$ metal poor dwarf galaxies  
\citep{Christensen2012, Stark2014a}.  In these systems, the FUV metal lines (CIV, CIII], OIII]) are typically much stronger 
than He II and NIV].   
The 2$\sigma$ upper limit on the He II to Ly$\alpha$ flux ratio ($<$0.07)  is 
nevertheless  consistent with the spectrum of BX 418 (f$_{\rm{HeII}}$/f$_{\rm{Ly\alpha}}$ = 0.03), a metal poor $z=2.3$ galaxy discussed in detail in \citet{Erb2010}.  A deeper J-band spectrum of A1703-zd6 
would be required to determine if He II is present at a similar flux as BX 418.

 \begin{figure*}
\begin{center}
\includegraphics[trim=0.15cm 0cm 1.0cm 0cm,width=1.0\textwidth]{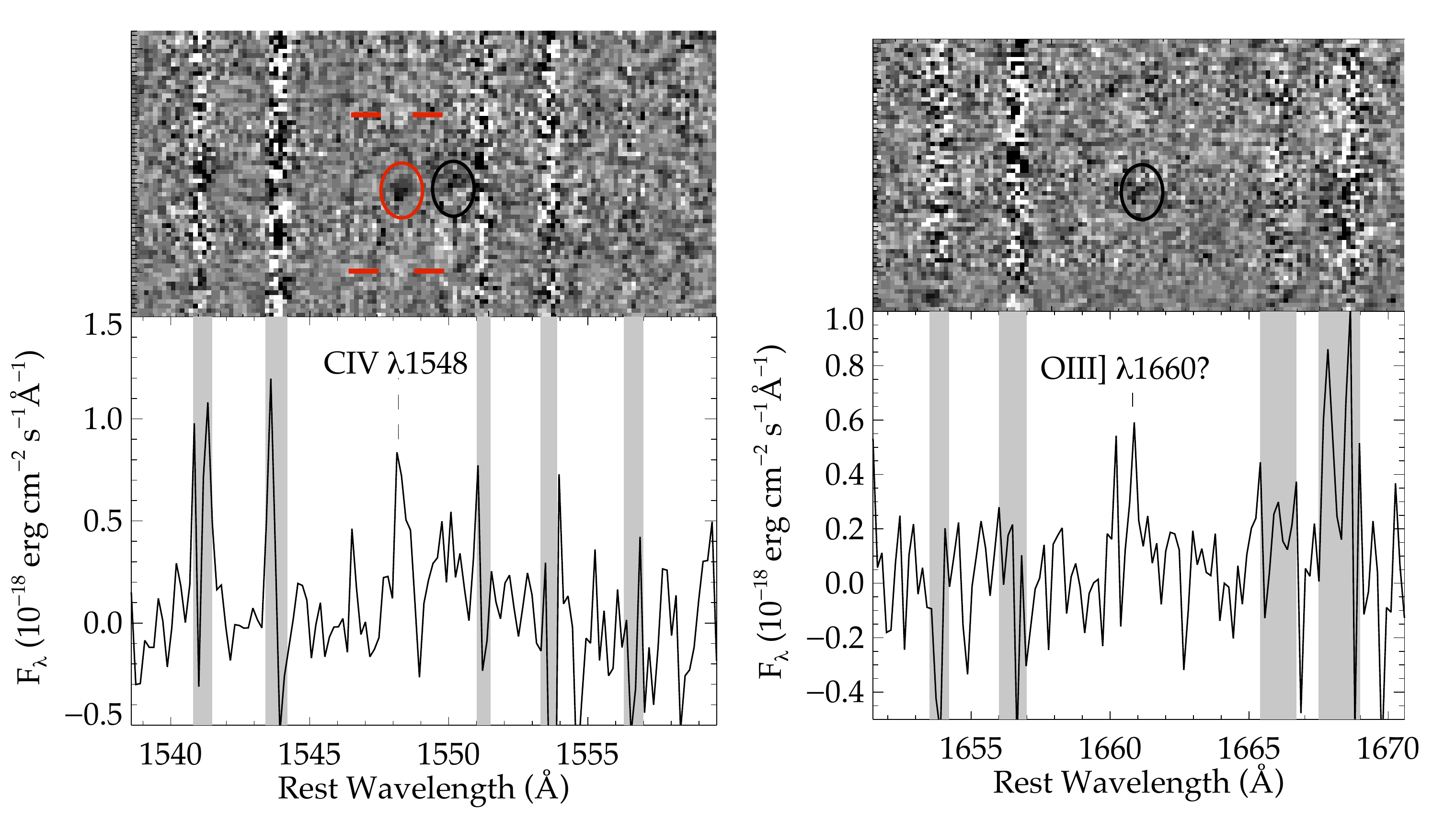}
\caption{Keck/MOSFIRE J-band spectrum of the gravitationally-lensed $z_{\rm{Ly\alpha}}=7.045$ galaxy A1703-zd6.  The spectroscopic redshift  was known prior to MOSFIRE observations from Ly$\alpha$ emission \citep{Schenker2012}.   Top panels show the two-dimensional (unsmoothed)  spectra with black showing positive emission.   The red oval identifies the emission feature at the rest-frame wavelength expected for CIV$\lambda$1548.   The 
characteristic negative emission from the dither pattern is demarcated by red horizontal lines.
Black ovals highlight the expected location of CIV$\lambda$1550 (left panel) and OIII]$\lambda$1660 (right panel).   The one-dimensional extracted 
spectra are shown below.   Vertical grey swaths indicate regions of elevated noise from OH sky lines.     }
\label{fig:spec}
\end{center}
\end{figure*}

\begin{table}
\begin{tabular}{lcccc}
\hline 
Line &  $\lambda_{rest}$  &  $\lambda_{obs}$  &  Line Flux  &  W$_0$  \\ 
         &  (\AA)                        &  (\AA)  & (10$^{-18}$  erg  cm$^{-2}$ s$^{-1}$) &    (\AA)    \\ \hline \hline
   Ly$\alpha^\dagger$   & 1215.67 &   9780 & $28.4\pm5.3$ & $65\pm12$  \\
   NIV]  &1483.3    &   \ldots & $<$3.6 & $<$15.7 \\
   \ldots  &  1486.5 & \ldots  & $<$4.3 &  $<$19.0 \\
   CIV &  1548.19 &  12457.9 & $4.1\pm0.8$ & $19.9\pm4.2$ \\
     \ldots  &  1550.77 &   12473.5 & $3.8\pm0.8$ & $18.1\pm4.3$ \\
   He II  & 1640.52 & \ldots  & $<$2.1&  $<$11.4 \\
   OIII] &  1660.81 & 13358.3 & $1.8\pm0.7$ &  $9.8\pm3.9$ \\
   \ldots &  1666.15 & \ldots  &  \ldots &  \ldots \\
   \end{tabular}
\caption{Emission line properties of the $z_{\rm{Ly\alpha}}=7.045$ galaxy A1703-zd6 (Bradley et al. 2011).    An aperture correction of 1.20$\times$ must  
be applied to the emission line fluxes if computing line luminosities.   The $\dagger$ next 
to Ly$\alpha$ notes that this flux measurement is from Schenker et al. (2012).    
The OIII]$\lambda$1666 emission line is obscured by a skyline.   The equivalent widths include the aperture 
correction and are quoted in the 
rest-frame.   The limits are 2$\sigma$. }
\end{table}
 
\subsection{A1703-zd1a}

A1703-zd1a is a very bright (H=23.9) z-band dropout.    After correcting for its lensing magnification ($\mu = 9.0 \pm 4.5$; Bradley et al. 2011), the absolute magnitude (M$_{\rm{UV}}=-$20.6) of A1703-zd1a  is found to be identical to L$^\star_{\rm{UV}}$ at $z\simeq 6.8$ \citep{Bouwens2014b}.   The UV slope ($\beta = -1.4$; \citealt{Smit2014a}) is  redder than average for L$^\star_{\rm{UV}}$ $z$-drops \citep{Bouwens2014a}.  Measurement of the  Spitzer/IRAC [3.6]-[4.5] color reveals a strong flux excess in [3.6], suggestive of strong [OIII]+H$\beta$ line contamination \citep{Smit2014a}.   This likely places the galaxy in the redshift range $z=6.6-6.9$ where [OIII]+H$\beta$ is in [3.6] and H$\alpha$ is between the [3.6] and [4.5] filters.   In this redshift 
range, the MOSFIRE J-band spectrum is sensitive to emission from CIV, 
He II, and OIII].   Since both components of CIV and OIII] are resolved, there are up to 
five emission lines which could be visible in the spectrum.  

Examination of the MOSFIRE spectrum reveals no features as strong as the CIV$\lambda$1548 line in A1703-zd6.   In regions between sky lines, the 5$\sigma$ upper limit to the line flux is 3.0$\times$10$^{-18}$ erg cm$^{-2}$ s$^{-1}$.   Given the seeing, slit width, and source size, we estimate an aperture 
correction of 1.27$\times$ is required for comparison of line fluxes to the total continuum 
flux densities.   Applying this correction factor to the line flux limit, we derive a 5$\sigma$ rest-frame 
equivalent width limit of 2.3~\AA, considerably lower than that seen in the spectrum of A1703-zd6.   
The spectrum is sensitive to CIV over $6.4<z<7.7$, covering the entire  redshift range suggested 
by the photometric redshift.

\subsection{A1703-zd4}

A1703-zd4 is  another bright (H$_{160}$=25.4) $z$-band dropout galaxy identified in 
\citet{Bradley2012}.   The broadband SED is best fit by a redshift of $z=8.4$ with acceptable solutions ranging between $z=7.0$ and $z=9.3$ \citep{Bradley2012}.   After taking into account the source magnification ($\mu=3.1$), the absolute magnitude of A1703-zd4 is $M_{\rm{UV}}=-20.6$ at 
its best fit photometric redshift.
 The J-band MOSFIRE spectrum is sensitive to Ly$\alpha$ over the redshift range $8.5<z<10.1$ and CIV over $6.4<z<7.7$.

We have visually examined the  spectrum for potential emission features.   While 
several low S/N features are present,  no definitive redshift identification is possible with 
the current spectrum.   Conservatively assuming a line width of 10~\AA\ (twice the value of the CIV$\lambda$1548 line in A1703-zd6) and 1\farcs1 aperture along the slit, we find 
that typical 5$\sigma$ line flux limits are 2.6$\times$10$^{-18}$ erg cm$^{-2}$ s$^{-1}$ in between OH sky lines.   Taking into account the seeing, slit 
width, and galaxy size, we compute an aperture correction of 1.21$\times$.   Applying this 
correction to the line  flux limit and using the J$_{125}$-band magnitude from \citet{Bradley2012} as the 
continuum flux density at 1.15-1.35$\mu$m, we derive a rest-frame equivalent width limit of  $\simeq 6$~\AA\ for regions 
between OH lines.  With the resolution provided by MOSFIRE, the incidence of OH 
lines is minimized, but nonetheless roughly 40\% of the J-band spectral window is still impacted by sky lines so it is of course possible that an emission line from A1703-zd4 is 
obscured by a bright skyline.   

Additional spectroscopy will help clarify the redshift of A1703-zd4, as the current J-band 
exposure only samples a portion of the photometric redshift distribution function.   A Y-band spectrum would extend the Ly$\alpha$ coverage down to the redshift range $7.0<z<8.2$, while an H-band spectrum of A1703-zd4 would extend the redshift range over which FUV metal lines are detectable.  

\subsection{A1703-23 }

A1703-23 was identified as a bright (H=23.75) $i$-band dropout and spectroscopically-confirmed in \citet{Richard2009}.     Ly$\alpha$ was detected in an optical spectrum 
at 8300.5~\AA\ implying a  redshift of $z=5.828$, consistent with expectations from 
the SED.   \citet{Richard2009}  measure a  Ly$\alpha$ flux of 2.5$\times$10$^{-17}$
erg cm$^{-2}$ s$^{-1}$.    After correcting for the magnification factor estimated by \citet{Richard2009}, $\mu=3$,  the UV absolute magnitude of A1703-23 is 
found to be M$_{\rm{UV}} = -21.7$, corresponding to a 1.6-2.0L$^\star_{\rm{UV}}$ galaxy at $z\simeq 5.9$ \citep{Bouwens2014b}.  Similar to A1703-zd1a, the UV spectral slope of A1703-23 ($\beta = -1.5$) is redder than typical $i$-band dropouts.    Since the continuum is not detected 
in the optical spectrum, we estimate the Ly$\alpha$ equivalent width from the broadband 
flux measurements.   We use the UV spectral slope to convert the J-band flux to a continuum 
flux at the wavelength of Ly$\alpha$.   Following this procedure, we estimate a rest-frame 
equivalent width of W$_{\rm{Ly\alpha}}$=10.3~\AA\ for A1703-23.

Using the Ly$\alpha$ spectroscopic redshift, we  predict the  
observed wavelengths over which the CIII] doublet will be located.   Studies at lower redshift  
demonstrate that CIII] tends to trace the systemic redshift (\citealt{Stark2014a}).   
Allowing for the characteristic Ly$\alpha$ velocity offset of 
0 to 450 km s$^{-1}$ with respect to the systemic redshift, we predict that 
[CIII]$\lambda$1907 will lie between 1.2999 and 1.3019 $\mu$m and 
CIII]$\lambda$1909 will be located between 1.3013 and 1.3033 $\mu$m.   The flux 
ratio of [CIII]$\lambda1907$ and CIII]$\lambda$1909 is set by the electron 
density in the gas traced by doubly-ionized 
carbon.   In high redshift galaxies, observations indicate [CIII]$\lambda$1907/CIII]1909 flux ratios in the range 1.2-1.6 \citep{Hainline2009,James2014}.  

As can be seen in the the 2D spectrum (Figure 3), no strong emission lines are 
detected.   The region over which [CIII]$\lambda$1907 is expected is almost entirely 
devoid of OH sky lines.   If we conservatively assume a FWHM of 10~\AA\ (roughly 2$\times$ as large as the unresolved CIV line in A1703-zd6), we find that the 
5$\sigma$ limit on the [CIII]$\lambda$1907 line flux would be 2.5$\times$10$^{-18}$ erg cm$^{-2}$ s$^{-1}$.    Roughly 70\% of the CIII]$\lambda$1909 spectral window is clean with a 5$\sigma$ flux limit of 5.6$\times$10$^{-18}$ erg cm$^{-2}$ s$^{-1}$.    The CIII]$\lambda$1909 
flux limit is slightly larger because the adopted 10~\AA\ line width overlaps with the OH line.   
 Applying an  aperture correction of 1.66$\times$ (calculated from the seeing, slit width, and source size), 
we derive 5$\sigma$ rest-frame equivalent width limits of 1.9~\AA\ for [CIII]$\lambda$1907 
and 4.3~\AA\ for CIII]$\lambda$1909.   Detection of both components of the doublet 
would require a total equivalent width in excess of 6.2~\AA. 

The non-detection of CIII] in A1703-23 is consistent with the physical picture presented 
in \citet{Stark2014a}.    At intermediate redshifts, the CIII]  equivalent 
width is found to increase with the Ly$\alpha$ equivalent width.   For galaxies with 
Ly$\alpha$ equivalent widths similar to the moderate value observed in A1703-23, 
CIII] equivalent widths are seen to be in the range W$_{\rm{CIII]}}\simeq 2$-4~\AA\ \citep{Shapley2003,Stark2014a}, below 
the equivalent width limit provided by the MOSFIRE spectrum.     However recent studies have 
demonstrated that some galaxies at $z\simeq 6-7$ have UV metal lines that are 
stronger than found in the intermediate redshift samples.    The non-detection of CIII] 
in A1703-23 suggests that  not all $z\gsim 6$ galaxies have such extreme rest-UV spectra.

\begin{figure}
\begin{center}
\includegraphics[width=0.48\textwidth]{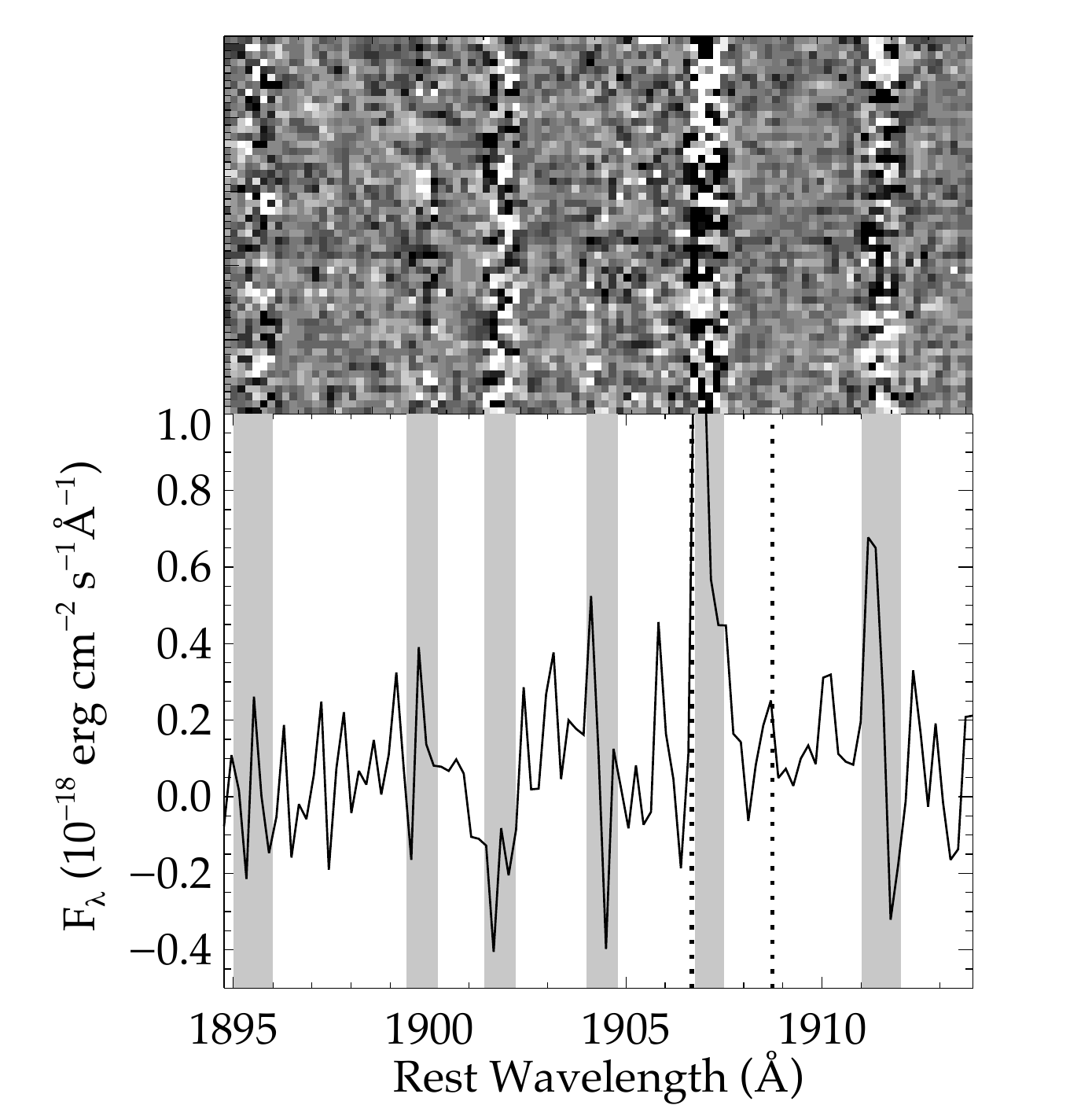}
\caption{Non-detection of [CIII]$\lambda$1907,CIII]$\lambda$1909 doublet in a 
Keck/MOSFIRE J-band spectrum of the spectroscopically-confirmed $z=5.828$ galaxy A1703-23 (see \S3.4 for details).    Top panel shows the two-dimensional (unsmoothed)  spectra with black corresponding to positive emission.   The one-dimensional extracted 
spectra is shown below.   Vertical grey swaths indicate regions of elevated noise from OH sky lines, and vertical dotted lines show expected location of CIII] doublet at the redshift of Ly$\alpha$.    }
\label{fig:spec}
\end{center}
\end{figure}

\section{A Hard Ionizing Spectrum at $z=7$}

Here we use photoionization models to characterize the shape of the 
ionizing spectrum required to produce the observed spectral properties of A1703-zd6.    We explore galaxy  
models with stellar input spectra in \S4.1 and AGN models in \S4.2. 

\subsection{Stellar models}
We fit the observed CIV$\lambda$1548 emission equivalent width (the component of the doublet that is most robustly 
detected) and broadband F125W and F160W fluxes of the galaxy using an approach similar to that adopted in \citet{Stark2014a}. This is based on a combination of the latest version of the \citet{Bruzual2003} stellar population synthesis model with the standard photoionization code CLOUDY \citep{Ferland2013} to describe the emission from stars and the interstellar gas (Gutkin et al., in preparation, who follow the prescription of \citealt{Charlot2001}). 

We adopt the same parameterization of interstellar gas and dust as in \citet{Stark2014b}, to whom we refer for detail. In brief, the main adjustable parameters of the photoionized gas are the interstellar metallicity, $Z$, the typical ionization parameter of a newly ionized HII region, $U$ (which characterizes the ratio of ionizing-photon to gas densities at the edge of the Str\"{o}mgren sphere), and the dust-to-metal (mass) ratio, $\xi_{\mathrm d}$ (which characterizes the depletion of metals on to dust grains). We consider here models with hydrogen densities ranging between 100 and $1000\,\mathrm{cm}^{-3}$ and C/O (and N/O) abundance ratios ranging from 1.0 to 0.1 times the standard values in nearby galaxies [$(\mathrm{C/O})_\odot\approx0.44$ and $(\mathrm{N/O})_\odot\approx0.07$]. We also include attenuation of line and continuum photons by dust in the neutral ISM, using the 2-component model of \cite{Charlot2000}, as implemented by \citet[][their equations~1--4]{Dacunha2008}. This is parameterized in terms of the total $V$-band attenuation optical depth of the dust, $\hat{\tau}_V$, and the fraction $\mu$ of this arising from dust in the diffuse ISM rather than in giant molecular clouds. Accounting for these two dust components is important to describe the different attenuation of emission-line and stellar continuum photons. We use a comprehensive model grid similar to that adopted by Stark et al. (2015) 
 
As in \citet{Stark2014b}, we consider models with 2-component star formation histories: a  `starburst' component (represented here by a 3\,Myr-old stellar population with constant SFR) and an `old' component (represented by a stellar population with constant or exponentially declining SFR with age between 10\,Myr and the age of the Universe at the galaxy redshift, i.e. 0.75\,Gyr). We adopt a standard \citet{Chabrier2003} initial mass function and the same stellar metallicity for both components, which also coincides with the current interstellar metallicity of the galaxy. 

To interpret the combined stellar and nebular emission from the galaxy, we use the same Bayesian approach as in Stark et al. (2015; see also equation~2.10 of \citealt{Pacifici2012}) and a  grid of models covering wide ranges in the above parameters. In practice, we find that the requirement to reproduce the strong CIV1548 emission equivalent width implies that the best-fit models are entirely dominated by the young stellar component (for reference, the best-fit model equivalent width for this line is 19.9\,\AA). Such models also provide excellent fits to the observed F125W and F160W fluxes (with a dust attenuation optical depth consistent with zero). Also, since nebular emission is constrained only by the equivalent width of a single component of the CIV doublet in our analysis, the resulting constraints on the gas density and C/O ratio are extremely weak.   

The models  demonstrate that the CIV emission line strength of A1703-zd6 can be reproduced by stellar input spectra.  The range of acceptable model parameters is shown in Table 3.
Models that fit the observed spectral properties  have a large  ionization parameter (log U = -1.35) and 
very low metallicity (12 + log O/H = 7.05).    The production rate of hydrogen ionizing photons 
per observed (i.e. attenuated) 1500~\AA\ luminosity is very large  ($\rm{log (\xi_{\rm{ion}}/erg^{-1}} Hz$)=25.68) in models 
that reproduce the data.   We note that we do not attenuate the ionizing photon 
output in our calculation of $\xi_{ion}$, as our primary goal is to be able to predict how many hydrogen 
ionizing photons were produced based on the observed 1500~\AA\ luminosity.

\begin{figure}
\begin{center}
\includegraphics[trim=1.25cm 5.5cm 1.5cm 3.25cm,width=0.48\textwidth]{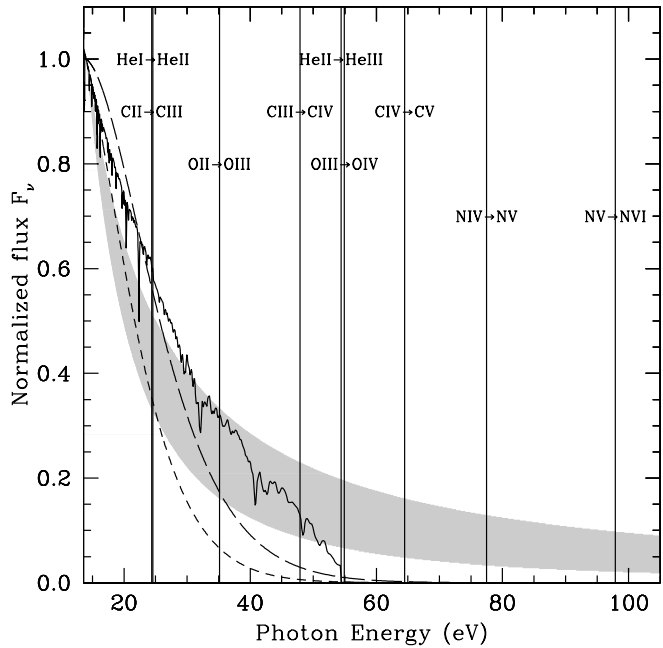}
\caption{Ionizing spectra from different sources, plotted together with the ionizing potentials of different ions (vertical lines). The solid spectral energy distribution corresponds to the best-fit galaxy model reported in Section \S4, while the shaded area shows the range in ionizing spectra produced by $F_{\nu}\propto\nu^{\alpha}$ AGN models with spectral indices between $\alpha=-1.2$ (upper ridge) and $-2.0$ (lower ridge). For comparison to the radiation field 
in typical galaxies at $z\simeq 2$ \citep{Steidel2014}, we also display blackbody spectra with temperatures of 45,000 (short dashed) and 55,000\,K (long dashed), respectively.     All spectra are normalized to unity at 13.6\,eV.}
\label{fig:model}
\end{center}
\end{figure}

The ionizing spectrum of the best-fitting stellar model (Figure \ref{fig:model}) reveals a significant flux of energetic radiation at 40-50 eV capable of producing 
nebular CIV emission.   The flux density drops off significantly above $\sim 54$ eV, resulting in much weaker 
emission from He II and NV.   Stellar models predict  NV$\lambda$1240 and He II$\lambda$1640 fluxes that 
are 30-250$\times$ weaker than the CIV$\lambda$1548 line strength (Table 3).  These lines are unlikely to be 
detected if the CIV emission is powered by a stellar population.

\begin{table}
\begin{tabular}{lr}
\hline 
\multicolumn{2}{c}{A1703-zd6 } \\
\hline\hline
                                      $\log{U}$   & $-1.35_{- 0.40}^{+ 0.24}$         \\
                      $12+\log{(\mathrm{O/H})}$   & $ 7.05_{- 0.25}^{+ 0.31}$         \\
    $\log(\xi_\mathrm{ion}/{\rm Hz\,erg}^{-1})$   & $25.68_{- 0.19}^{+ 0.27}$         \\
                         $W(\mathrm{Ly\alpha})$   & $ 2.06_{- 0.24}^{+ 0.33}$         \\
           $\log(\mathrm{NV\,1240/CIV\,1548})$   & $-2.43_{- 0.42}^{+ 0.42}$         \\
         $\log(\mathrm{HeII\,1640/CIV\,1548})$   & $-1.59_{- 0.20}^{+ 0.22}$         \\
         $\log(\mathrm{MgII\,2798/CIV\,1548})$   & $-1.95_{- 0.47}^{+ 0.42}$         \\
\end{tabular}
\caption{A1703-zd6 photoionization modelling results.   The  input 
spectra to the photoionization code are from the latest version of the 
\citet{Bruzual2003} stellar population synthesis models.    We fit the CIV$\lambda$1548 
equivalent width and J$_{125}$ and H$_{160}$ broadband flux densities.  
     }
\end{table}

\subsection{AGN models}

We also explore a comparison of A1703-zd6 to AGN photoionization models.   We use the models of Feltre et al. 2015 in prep., which represent the narrow line emission region (NLR) of the AGN and have been realized with the photoionization code CLOUDY \citep[latest version C13.03, last
described in][]{Ferland2013}.  The solar abundance and dust depletion values are the same as those used in the models for star-forming galaxies used in \S4.1 and developed 
by Gutkin et al. 2015, in prep.   The input parameters for the AGN NLR models are taken to be the interstellar metallicity, Z, in the range between 0.0001 and 0.05, the dust-to-metal(mass) ratio, $\xi_{\rm d}=0.1, 0.3$ and 0.5, the hydrogen density, n$_{\rm H}=10^{2}, 10^{3}$ and $10^{4}$ cm$^{-3}$, and the ionization parameter, ${\rm U}$, in the range  $\rm{log~U}$ =  $-5$ to $-1$.   The AGN is characterised by a power law  F$_{\nu} \propto \nu^{\alpha}$, with  spectral indices $\alpha=-1,2,-1.4,-1.7$ and $-2.0$ in the wavelength range between 0.001 and 0.25 $\mu$m.  For completeness, we include the effects of attenuation using the prescription of \cite{Calzetti2000}.

In contrast to the stellar model fits (\S4.1), we do not include the constraints from the line equivalent widths and 
broadband photometry because of the arbitrary scaling of the AGN and host galaxy luminosities that would be 
required to model the underlying continuum.   The AGN fit is instead focused on the two most robustly constrained  flux ratios, 
namely O{\sc iii}](1660)/C{\sc iv}(1548) and He{\sc ii}(1640)/C{\sc iv}(1548).   Given the tentative nature of the OIII]$\lambda$1660 detection, we 
adopt a 5$\sigma$ upper limit ($<$3.5$\times$10$^{-18}$ erg cm$^{-2}$ s$^{-1}$) to the line flux.   
We note that this limit is consistent with the observed 
faint flux level of the possible emission feature (1.8$\times$10$^{-18}$ erg cm$^{-2}$ s$^{-1}$).   
The measurement of C{\sc iv}$\lambda$1548 and upper limits on O{\sc iii}]$\lambda$1660 and He{\sc ii} thus translate into upper limits for the emission line ratios O{\sc iii}]$\lambda$1660/C{\sc iv}$\lambda$1548 and He{\sc ii}/C{\sc iv}$\lambda$1548 used in the fit. 

The fitting procedure demonstrates that the existing observational constraints are also marginally consistent with photoionization by an AGN.    The two  flux ratio limits favour best-fit AGN models corresponding to metallicity and hydrogen density of  Z=0.001 and 10$^{2}$cm$^{-3}$, respectively.   The ionizing spectra of 
acceptable AGN models is  shown in Figure 4.   The AGN power law spectrum has  
considerably greater flux than the galaxy models at  energies in excess of 50 eV.   As a result, AGN 
models predict stronger  
He II$\lambda$1640 than the  galaxy models discussed in \S4.2, with a flux that is 40\% that of the 
measured CIV$\lambda$1548 flux.   Current limits suggest the He II flux is less than 50\% that of CIV$\lambda$1548 at 2$\sigma$.   Deeper J-band observations may thus be able to detect He II if A1703-zd6 
is powered by an AGN.

\section{Discussion}

 \begin{figure}
\begin{center}
\includegraphics[trim=1cm 0.25cm 0.5cm 0cm,width=0.48\textwidth]{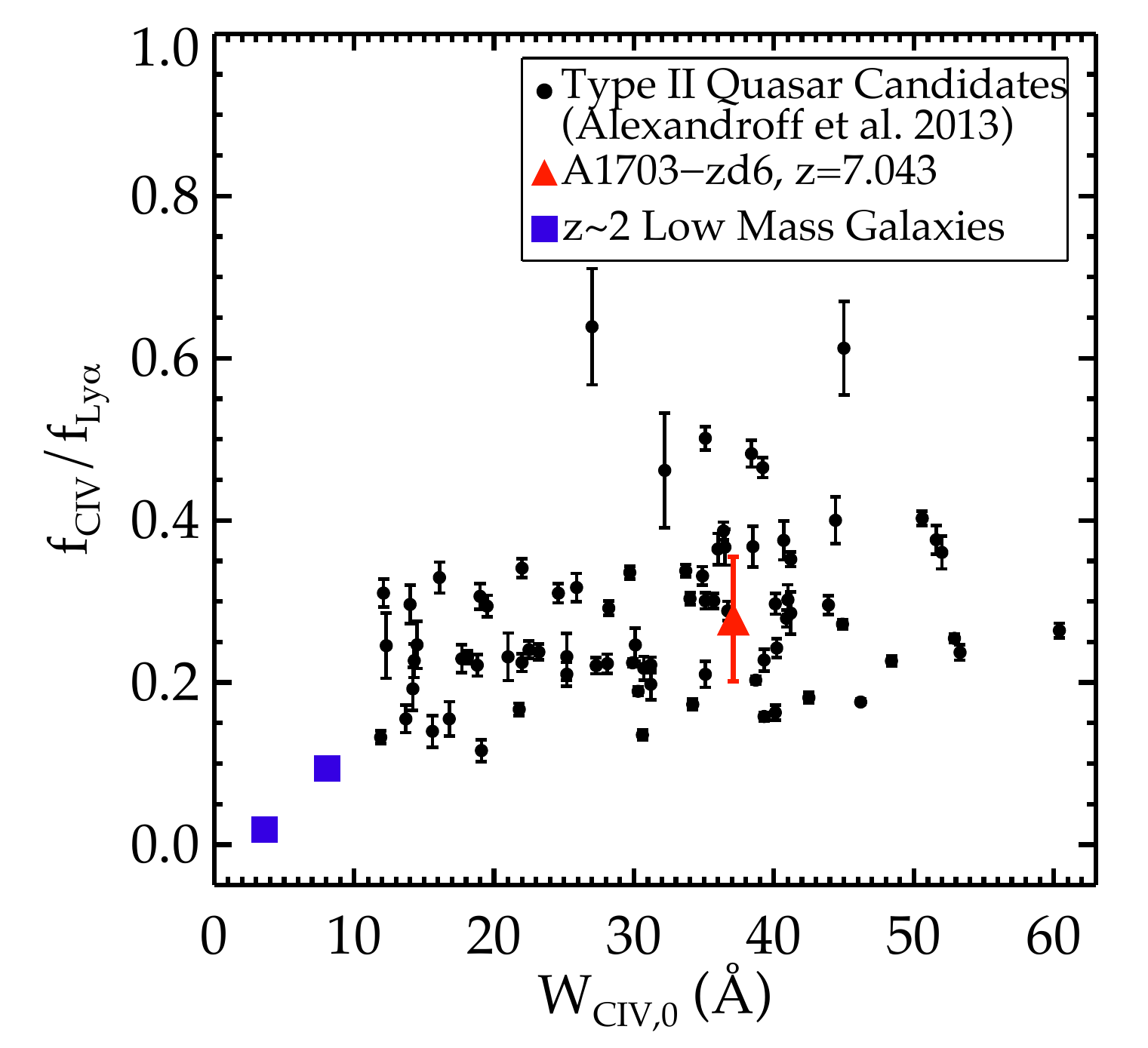}
\caption{Comparison of the CIV/Ly$\alpha$ flux ratio in A1703-zd6 with type II quasar candidates at $2<z<4.3$
and low mass metal poor galaxies at $z\simeq 2$.    The quasar candidates were reported in \citet{Alexandroff2013} and the low mass $z\simeq 2$ galaxies are CIV emitters from 
\citet{Stark2014a}.    The CIV emitting strength of A1703-zd6 is comparable to that seen in AGN 
spectra and is more extreme than in powerful line emitting galaxies at lower redshift.}
\label{fig:CIV}
\end{center}
\end{figure}

\subsection{Comparison to UV Radiation Field at Lower Redshifts}

Nebular CIV emission is seldom detected in $z\simeq 2-3$ galaxies.    Instead the 
CIV profile shows strong absorption from highly ionized outflowing gas  superimposed on a 
P-Cygni profile from stellar winds (e.g, \citealt{Shapley2003}).  The absence of strong CIV emission 
in typical galaxies points to a negligible output of photons with energies greater than 47.9 eV, consistent with 
the radiation field expected from  the blackbody models predicted for $z\simeq 2$ galaxies in \citet{Steidel2014}.  
In Figure \ref{fig:model}, we overlay the $z\simeq 2$ galaxy ionizing spectrum on the 
AGN and stellar photoionization models which  reproduce the spectral features of A1703-zd6.  
Both models predict that the $z=7.045$ galaxy must have a larger output of  20-50 eV 
radiation than is commonly seen in $z\simeq 2-3$ galaxies.  

When strong CIV emission is present in lower redshift systems, it is generally thought to 
reflect AGN activity.   At $z\simeq 2-3$, AGNs with narrow emission lines (FWHM less than 2000 km s$^{-1}$)
comprise only 1\% of UV-selected galaxy samples  \citep{Steidel2002,Hainline2011}.   In these systems, 
the CIV flux is typically 20\% that of Ly$\alpha$, with equivalent widths ranging between 10 and 50~\AA\ 
\citep{Steidel2002,Hainline2011,Alexandroff2013}, comparable to what is observed in A1703-zd6 (Figure \ref{fig:CIV}).  
Nebular He II is often seen in UV-selected narrow-lined AGNs, but the equivalent width  is  only 
5-10~\AA\  \citep{Hainline2011}, also consistent with the non-detection of He II in  A1703-zd6.    While 
the line properties of A1703-zd6 and UV-selected AGNs appear similar, the continuum shapes are very different.
 \citet{Hainline2011} demonstrated that at lower redshift,  the UV continuum power-law slope of the UV-selected AGNs ($\beta = -0.3$) tends to be much 
redder than non-AGNs ($\beta=-1.5$).   In contrast, A1703-zd6 has a very blue ($\beta=-2.4$)
continuum slope  \citep{Bradley2012,Schenker2012}, similar to the parent population of galaxies at $z\simeq 7$ (e.g., \citealt{Bouwens2014a}). 

Among star forming galaxies with  low stellar masses  (10$^6$-10$^9$ M$_\odot$), nebular CIV emission is 
more common  \citep{Christensen2012,Stark2014a}.    These systems tend to have blue colours, large specific 
star formation rates, and prominent Ly$\alpha$ emission lines.   The strong CIV emission is thought to reflect an intense radiation field from massive stars following a 
recent upturn in star formation \citep{Stark2014a}.   Given the similarly low masses, blue colours, and large 
specific star formation rates of reionization-era galaxies, it is possible that nebular CIV emission might 
be more common in $z>6$ galaxies.   Yet as is evident in Figure \ref{fig:CIV}, the CIV equivalent widths and 
CIV/Ly$\alpha$ flux ratios of A1703-zd6 are  5-10$\times$ larger than the low mass samples at lower redshift.   We note that
the reduced CIV/Ly$\alpha$ flux ratio of A1703-zd6 with respect to $z\simeq 2$ systems could be in part due to 
suppression of Ly$\alpha$ emission by the IGM or optically thick absorbers.   But given the very large rest-frame 
Ly$\alpha$ equivalent width of this system (65~\AA, \citealt{Schenker2012}),  IGM attenuation is not likely 
to be the primary factor responsible for the nearly order of magnitude offset in the CIV/Ly$\alpha$ ratio compared to the lower redshift systems.

There does not appear to be a completely analogous population to A1703-zd6 at lower redshift.   Deeper spectroscopic 
constraints on other ultraviolet spectral features (NV, He II, CIII]) will help characterize the powering mechanism.  
Another crucial step will be determining whether  systems like A1703-zd6 are common at $z\simeq 7$.   
The only other galaxy  in our sample with a redshift which places CIV in the spectral window we targeted is A1703-zd1a (\S3.2).    The non-detection of 
CIV in this system  suggests that the line is not strong in all $z\simeq 7$ galaxies.    But  
presence of nebular CIV emission in one of two galaxies we observed represents a significant 
departure from the 1\% detection rate at lower redshift.   If the galaxies in our sample are 
typical of the UV-selected population at $z\simeq 7$, it would suggest a very different 
UV radiation field is present in many early star forming systems, potentially altering our 
current picture of the contribution of galaxies to reionization (e.g., Robertson et al. 2015, Bouwens 
et al. 2015).   If the duty cycle associated with the 
intense radiation field is large (such that A1703-zd6 is not a bursty outlier), then it would be 
possible to achieve reionization by $z\simeq 6$ while adopting a lower ionizing photon escape 
fraction (f$_{\rm{esc}}$) or a brighter absolute magnitude limit when integrating the 
UV luminosity function.

\subsection{Nature of $z>7$ Ly$\alpha$ emitters}

Currently all $z>7$ UV metal line detections  come from  galaxies with 
previously known Ly$\alpha$ detections.    This  largely reflects our selection criteria, as we have 
primarily focused our follow-up on known Ly$\alpha$ emitters.  
The spectroscopic redshifts provided by Ly$\alpha$ are very useful, allowing us to ensure that UV 
metal lines are located at wavelengths which are 
unobscured by the atmosphere.     But given the unusually strong CIV emission, we must consider whether  our pre-selection of Ly$\alpha$ emitters 
might have biased us toward locating  galaxies with extreme radiation fields.  

While Ly$\alpha$ emission is very common among $z\simeq 6$ galaxies \citep{Stark2011}, 
it is exceedingly rare at $z\simeq 7$ 
(e.g., \citealt{Treu2013,Schenker2014, Tilvi2014}).   
The weak Ly$\alpha$ emission is thought to reflect an increase in the IGM neutral hydrogen fraction over 
$6<z<8$ \citep{Mesinger2014,Choudhury2014}, as would be expected if reionization is 
incomplete at $z\simeq 7-8$ (e.g, \citealt{Robertson2013, Robertson2015}).    Recent work suggests 
that an increased incidence of optically thick absorbers may also contribute to the 
attenuation of Ly$\alpha$ \citep{Bolton2013}, although the importance of such systems remains 
unclear \citep{Mesinger2014}.
Based on this overall picture, galaxies with Ly$\alpha$ detections at $z>7$, such as A1703-zd6,
will be those that are situated in regions of the IGM which have already been ionized.      

It is conceivable that the only $z>7$ galaxies that can be seen in Ly$\alpha$ are those with 
extreme radiation fields capable of ionizing   hydrogen in the IGM and in 
optically thick absorbers.  In this case, the presence of high ionization features like CIV in 
A1703-zd6 would be a direct consequence of 
the pre-selection by Ly$\alpha$.   Given the rarity of Ly$\alpha$ emitters at $z>7$, the hard 
ionizing spectrum shown in Figure 4 would thus be limited to a small percentage of the early 
galaxy population.    Additional constraints on the strength of high ionization lines in $z>7$ Ly$\alpha$ emitters will be needed 
to determine whether the presence of Ly$\alpha$ is indeed linked to the intensity of the 
radiation field.   Likewise, determination of the typical ionizing spectrum will require sampling 
galaxies with a diverse range of properties, including those without Ly$\alpha$ emission.

\subsection{Implications for early galaxy spectroscopy}

\begin{figure}
\begin{center}
\includegraphics[width=0.48\textwidth]{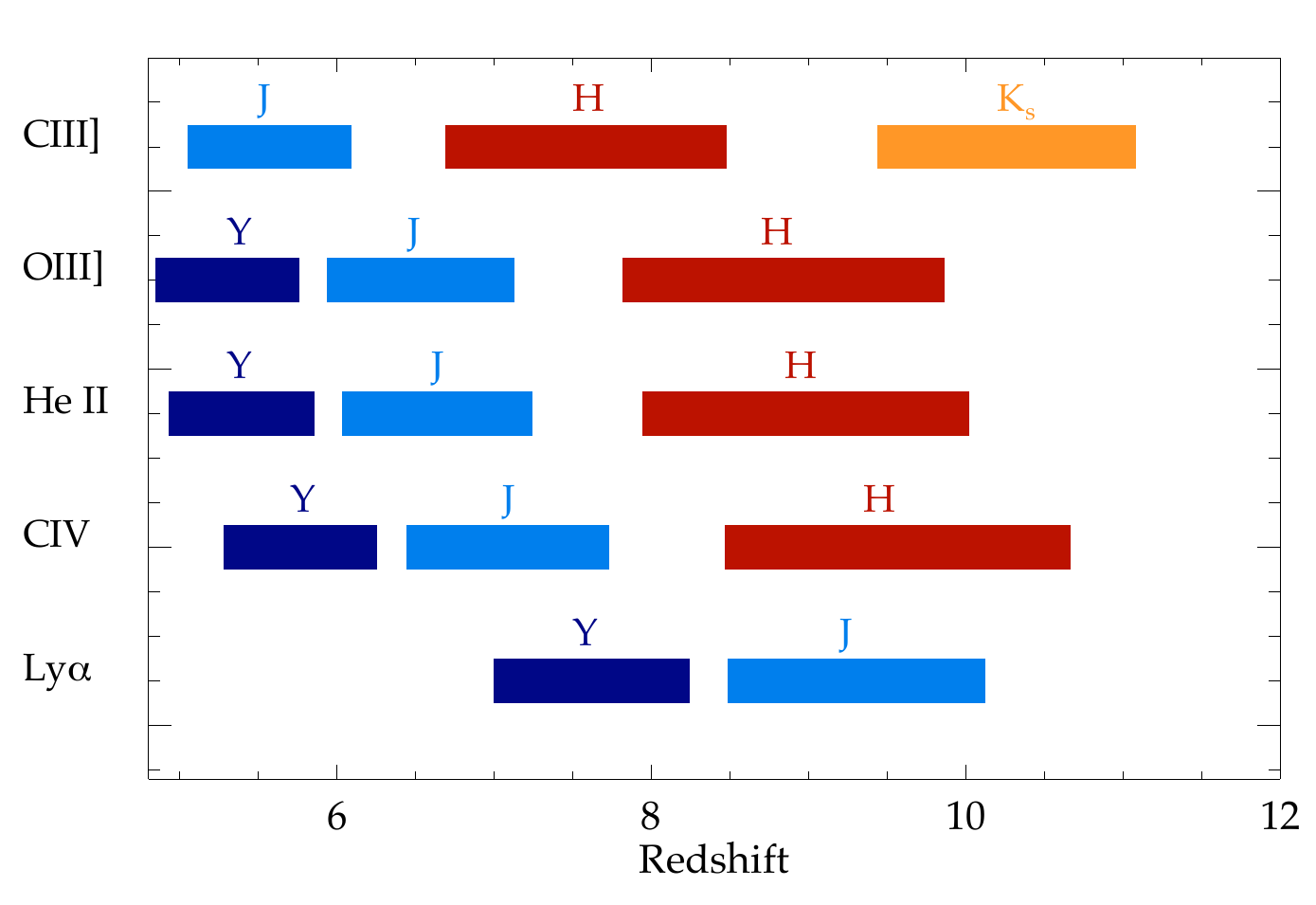}
\caption{Visibility of FUV lines in $z>5$ galaxies using ground-based spectrographs.   Solid bands give 
redshift window over which emission lines are detectable in Y, J, H, and K$_s$-band filters using MOSFIRE 
on Keck.    
Atmospheric absorption in the 
near-infrared limits the redshift range over which any individual spectral feature can be detected.     }
\label{fig:fuv}
\end{center}
\end{figure}

Recent studies have proposed that the [CIII]$\lambda$1907, CIII]$\lambda$1909 doublet 
provides a feasible route toward spectroscopic confirmation of reionization-era galaxies in which Ly$\alpha$ is strongly attenuated by the partially neutral IGM \citep{Erb2010,Stark2014a}.   The line may provide a valuable 
spectroscopic probe of the most distant galaxies that the James Webb Space Telescope (JWST) will find, as the 
strong rest-optical lines shift out of the spectral window of NIRSPEC at $z\simeq 11$.   For future ground-based 
optical/infrared telescopes, CIII] may be brightest spectral lines in reionization-era systems.   
 But because of  atmospheric absorption in 
the near-infrared, ground-based facilities will not be able to detect CIII] emission throughout substantial 
redshift intervals in the reionization era.   The low atmospheric transmission between the J and H-band and H and 
K$_s$-band causes CIII] to be attenuated in galaxies at  $5.8\lsim z\lsim 6.9$ and $8.3\lsim z\lsim 10$, respectively (Figure \ref{fig:fuv}).    

The presence of CIV and tentative identification of OIII] in A1703-zd6 suggests that a variety of FUV lines 
may be present in deep near-infrared spectra of a subset of bright $z>6$ galaxies.   
While it is possible that A1703-zd6 may have atypically strong CIV emission (see \S5.2), 
both CIV and OIII] emission are commonly detected in the spectra of intermediate redshift 
metal poor galaxies \citep{Erb2010,Christensen2012,Stark2014a}.  Importantly, both lines are detectable 
at redshifts where CIII] is unobservable.  CIV can be seen in galaxies 
over $6.6<z<7.4$ (J-band), and $8.7<z<10.5$ (H-band), while OIII] can be identified in galaxies 
at $6.0<z<6.8$ and $8.0<z<9.7$  (Figure \ref{fig:fuv}). 

The redshift-dependent visibility of the  FUV metal lines shown in Figure \ref{fig:fuv} has implications for the optimal 
strategy of spectroscopically following up early star forming systems.  
Of particular interest is the population of extreme [OIII]+H$\beta$ emitting galaxies identified via  blue [3.6]-[4.5] 
colours  \citep{Smit2014a,Smit2014b} and thought to lie in the redshift range $z\simeq 6.6-6.9$.   \citet{Stark2014a} demonstrate that  galaxies with extreme optical line emission tend  to have large equivalent width FUV metal lines 
(see Gutkin et al. 2015, in preparation for more details), making this an attractive class of objects to target with 
rest-frame FUV spectroscopy.  However Figure \ref{fig:fuv} demonstrates that CIII] is 
likely to be obscured by the atmosphere over some of  this redshift range.    For the same reason, attempts to identify CIII] emission in 
 Ly$\alpha$ emitters at $z=6.6$ will not result in many detections.  
Follow-up efforts of both populations are more likely to succeed by focusing on CIV, He II, and OIII] in the J-band.   

\section{Summary}

We have presented new measurements of various far-ultraviolet emission lines (Ly$\alpha$, CIV, He II, OIII]) in 
$z\sim 6-8$ galaxy spectra.   Using the MOSFIRE spectrograph \citep{Mclean2012} on Keck I, we have 
obtained J-band spectra of four gravitationally lensed galaxies in the massive cluster field 
Abell 1703 (Table 1).    The galaxies are bright (H=23.8-25.9) and span a  range of 
redshifts ($5.8\lsim z\lsim 8.0$), Ly$\alpha$ equivalent widths, UV slopes, and 
continuum luminosities.    

We report a 5.1$\sigma$ detection of nebular CIV$\lambda$1548 emission in A1703-zd6, a 
$z=7.045$ spectroscopically-confirmed galaxy.   The observed wavelengths of the UV metal lines are accurately 
known from the Ly$\alpha$ redshift.  
The rest-frame CIV equivalent width (38.0~\AA) and CIV/Ly$\alpha$ flux ratio (0.3) we infer for A1703-zd6 are similar to those 
seen in narrow-lined AGNs at lower redshift.   The line is unresolved with a FWHM less than 125 km 
s$^{-1}$.   The presence of CIV  requires a surprisingly large  
output of very energetic radiation capable of triply ionizing carbon.   In \S4, we demonstrated that the extreme 
spectral properties of A1703-zd6 can be powered by an AGN or an intense population of young, very hot, metal 
poor stars.   Photoionization models point to an ionizing spectrum that is very different from  that inferred 
for typical $z\simeq 2$ galaxies.

The data provide constraints on UV metal line emission in two other galaxies.    The first of these is 
A1703-zd1a, a $z$-band dropout thought to lie at $z\simeq 6.6-6.9$ \citep{Smit2014a}.   In this redshift 
range, the MOSFIRE observations probe CIV, He II, and OIII].   No secure 
emission lines  are identified throughout the J-band spectrum.  Rest-frame equivalent widths 
greater than 2.3~\AA\ would have been seen at 5$\sigma$ if located between OH sky lines.   
The other system is A1703-23, a spectroscopically confirmed galaxy at $z=5.828$.    We report an upper 
limit of 1.9~\AA\ on [CIII]$\lambda$1907 and 4.3~\AA\ on CIII]$\lambda$1909 emission.    The lack of 
detectable CIII] emission is consistent with the relationship between CIII] and Ly$\alpha$ equivalent 
width presented in \citet{Stark2014a}.

At lower redshift, nebular CIV emission is seen in only 1\% of UV-selected galaxies 
\citep{Steidel2002,Hainline2011}.    The presence of strong CIV emission in A1703-zd6, one 
of the first galaxies we targeted is thus rather surprising.  One explanation is that 
radiation fields are more extreme in $z\simeq 7$ galaxies.   This may be expected if reionization-era 
galaxies are commonly caught following a recent upturn or burst of star formation and would indicate 
that $z>7$ systems are more efficient ionizing agents than we previously thought.    Alternatively, we 
consider whether the only systems   seen in Ly$\alpha$ at $z>7$ may be those with 
extreme radiation fields capable of ionizing hydrogen in optically thick absorbers and the surrounding IGM.    In this case,  the pre-selection of galaxies by their Ly$\alpha$ emission would result in a very biased 
sample  with extreme spectral features.   Further  constraints on high ionization 
emission lines in  galaxies with and without Ly$\alpha$ should clarify whether systems 
like A1703-zd6 are ubiquitous at $z>7$.

\section*{Acknowledgments}
We thank Richard Ellis, Dawn Erb, Martin Haehnelt, Juna Kollmeier, Ian McGreer, and 
Andrei Mesinger for enlightening conversations.   DPS acknowledges support from the 
National Science Foundation  through the grant AST-1410155.   JR acknowledges support 
from the European Research Council (ERC) starting grant CALENDS and the Marie Curie Career Integration Grant 294074.   
SC, JG and AW acknowledge support from the ERC via an Advanced Grant under grant agreement no. 321323 -- NEOGAL.  
This work was partially supported  by a NASA Keck PI Data Award, administered by the 
NASA Exoplanet Science Institute. Data presented herein were obtained at the W. M. Keck Observatory 
from telescope time allocated to the National Aeronautics and Space Administration through the agency's 
scientific partnership with the California Institute of Technology and the University of California. 
The Observatory was made possible by the generous financial support of the W. M. Keck Foundation.
The authors acknowledge the very significant cultural role that the
summit of Mauna Kea has always had within the indigenous Hawaiian community.
We are most fortunate to have the opportunity to conduct observations from this mountain.

\bibliographystyle{mn2e}
\bibliography{references}

\label{lastpage}

\end{document}